\documentclass{article}
\usepackage{spconf,amsmath,graphicx}
\usepackage{url}
\usepackage[export]{adjustbox}
\usepackage{amssymb}
\usepackage{color,soul}

\renewenvironment{thebibliography}[1]{
  \begin{oldthebibliography}{#1}
    \setlength{\itemsep}{0.65em}
    \setlength{\parskip}{0em}
}
{
  \end{oldthebibliography}
}

\usepackage{enumitem}
\setlist{nosep, leftmargin=14pt}

\usepackage{mwe} 
\usepackage{tabularx}
\usepackage{booktabs}

\title{Single volume lung biomechanics from chest computed tomography using a mode preserving generative adversarial network}
\makeatletter
\def\@name{{\em Muhammad F. A. Chaudhary}$^{\star}$\quad {\em Sarah E. Gerard}$^{\star}$\quad {\em Di Wang}$^{\star}$\quad {\em Gary E. Christensen}$^{\star}$\\*[3pt]
  {\em Christopher B. Cooper}$^{\dagger}$\quad {\em Joyce D. Schroeder}$^{\ddagger}$\quad {\em Eric A. Hoffman}$^{\star}$\quad {\em Joseph M. Reinhardt}$^{\star}$\\}
\makeatother

\address{$^{\star}$ University of Iowa, Iowa City, IA, USA \qquad $^{\dagger}$ University of California, Los Angeles, CA, USA \\ $^{\ddagger}$ University of Utah, Salt Lake City, UT, USA}

\begin{document}
%

\maketitle
\begin{abstract}
Local tissue expansion of the lungs is typically derived by registering computed tomography (CT) scans acquired at multiple lung volumes. However, acquiring multiple scans incurs increased radiation dose, time, and cost, and may not be possible in many cases, thus restricting the applicability of registration-based biomechanics. We propose a generative adversarial learning approach for estimating local tissue expansion directly from a single CT scan. The proposed framework was trained and evaluated on $2500$ subjects from the SPIROMICS cohort. Once trained, the framework can be used as a registration-free method for predicting local tissue expansion. We evaluated model performance across varying degrees of disease severity and compared its performance with two image-to-image translation frameworks -- UNet and Pix2Pix. Our model achieved an overall PSNR of $18.95$ decibels, SSIM of $0.840$, and Spearman’s correlation of $0.61$ at a high spatial resolution of $1$mm$^3$.
\end{abstract}
\section{Introduction}
\label{sec:intro}

Computed tomography (CT) is a standard tool for diagnosing pulmonary diseases such as chronic obstructive pulmonary disease (COPD), lung cancer, and cystic fibrosis. Advancements in CT hardware have enabled high-resolution chest scans that are able to resolve subtle structural details of the lung tissue, airways, and vasculature. While CT reveals underlying structure, it fails to explicitly capture the lung function which is critical for understanding the extent of disease. 

\begin{figure}[ht]
\centering
\includegraphics[scale = 1, center, trim={0.7cm 0 1.35cm 0}, clip]{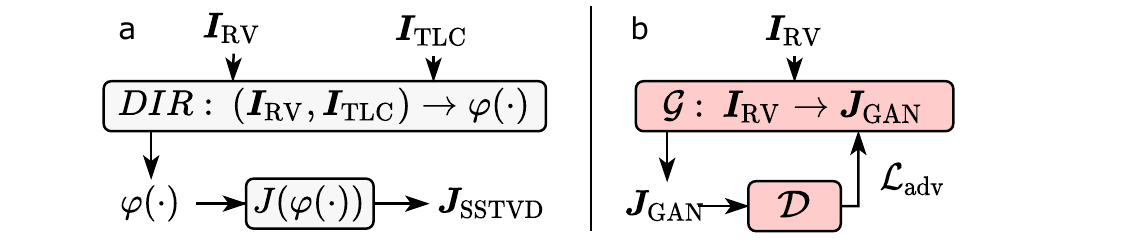}
\caption{(a) Image registration-based estimation of local tissue expansion. (b) The proposed single-volume, generative adversarial approach for estimating tissue volume change, trained such that $\boldsymbol{J}_{\mathrm{SSTVD}} \approx \boldsymbol{J}_{\mathrm{GAN}} $. $J(\cdot)$ operator refers to Jacobian determinant of the transformation field $\varphi(\cdot)$.}
\label{conv_to_dl}
\end{figure}

Deformable image registration (DIR) is used to find the deformation field which aligns multiple CT scans acquired at different volumes or time points. The determinant of the Jacobian matrix, referred to simply as the Jacobian, of the registration displacement field is used to quantify local tissue volume change between lung volumes. Registration between inspiratory-expiratory volume pairs have been analyzed to compute biomechanical properties of the lung~\cite{reinhardt2008registration}.  Xenon-enhanced CT (Xe-CT) was used to validate the registration-derived local tissue expansion as a surrogate for lung ventilation~\cite{reinhardt2008registration}. Furthermore, local tissue expansion between inspiration and expiration scans was shown to be associated with decline in lung function~\cite{bhatt2017computed}, and is also being used as ventilation surrogate for functional avoidance in radiation therapy for lung cancer~\cite{patton2018quantifying}. 

One approach for computing regional measures of lung function requires registration between two CT scans acquired at different lung volumes. However, acquiring CT scans at standardized lung volumes (full inspiration and full expiration) is clinically challenging in  patients who have trouble reliably performing multiple breath-hold scans due to reduced lung function. Additionally, acquiring multiple CT scans increases radiation dose, time, and cost. These problems limit the use of registration-derived biomechanical measures to patients with multiple scans. Furthermore, registration between high-resolution 3D CT images has a high computational cost. These limitations create a need for faster, more robust, and scalable methods for estimating lung function directly from single volume CT scans.

We hypothesized that a generative adversarial network (GAN) can be used to estimate local tissue expansion, between expiration and inspiration, directly from a single expiratory volume scan, assuming that a single image contains sufficient information for predicting regional mechanics. Therefore, we proposed to train a generative framework of two networks that compete against each other until they reach Nash's equilibrium~\cite{goodfellow2014generative} and are able to synthesize realistic estimates of Jacobian images.

\section{Methods}
\label{sec:methods}

\subsection{Dataset}
We analyzed data from SubPopulations and Intermediate Outcome Measures in COPD Study (SPIROMICS), which is an on-going, multi-center prospective cohort study aimed at identifying novel sub-types of COPD~\cite{couper2014design}. SPIROMICS acquired breath-hold CT scans at two lung volumes -- total lung capacity (TLC) and residual volume (RV). The study used a standardized CT image acquisition protocol and gathered data at $14$ different clinical sites across the US. The original resolution of the CT scans was approximately $0.6\times0.6\times0.5$ mm$^3$. Each slice consisted of $512\times512$ voxels and each image had $500$ to $600$ slices. We used CT data from $2500$ subjects at baseline with varying degrees of disease severity graded by the Global Initiative for Chronic Obstructive Lung Disease (GOLD) system -- GOLD 1 (mild) to GOLD 4 (severe)~\cite{vestbo2013global}. In this classification system, asymptomatic smokers were grouped in GOLD 0 and individuals who never smoked were treated separately (see Table \ref{t_1}).

\begin{table}[t!]
    \centering
    \caption{Distribution of COPD severity, defined by GOLD~\cite{vestbo2013global}, across disjoint training and testing datasets.}
    \vspace{0.1in}
    \begin{tabularx}{0.33\textwidth}{l c c}
    \toprule
      & Training & Testing \\
    \midrule
    GOLD 0     & $537$ & $225$ \\
    GOLD 1     & $268$ & $113$ \\
    GOLD 2     & $454$ & $189$ \\
    GOLD 3     & $287$ & $122$ \\
    GOLD 4     & $85$ & $37$ \\
    Non-smokers (N)  & $130$ & $53$ \\
    \midrule
    Total        & $1761$ & $739$ \\
    \bottomrule
\end{tabularx}
    \label{t_1}
\end{table}

\begin{table}[t!]
    \centering
    \caption{Quantitative evaluation of the proposed framework on $130,516$ slices from $739$ test subjects. Mann Whitney's U test was performed to assess differences between quantitative metrics of the proposed and other models, where `$\star\star$' indicated $p < 0.0001$ and `$\star$' indicated $p < 0.01$.}
    
    \vspace{0.1in}
    \setlength{\tabcolsep}{1.5pt}
    \begin{tabularx}{0.48\textwidth}{l l l l l l}
    \toprule
      & PSNR (dB) & SSIM & DSC$_{\mathrm{high}}$ & $r_{\mathrm{s}}$ & MAE \\
    \midrule
    UNet$_{\mathrm{SSIM}}$     & $18.44^{\star\star}$ & $0.832^{\star\star}$ & $0.918^{\mathrm{ns}}$ & $0.60^{\star\star}$ & $0.213^{\star\star}$ \\
    Pix2Pix~\cite{isola2017image}  & $18.13^{\star\star}$ & $0.825^{\star\star}$ & $0.915^{\star\star}$ & $0.57^{\star\star}$ & $0.218^{\star\star}$ \\
    \midrule
    Ours  & $18.90^{\star}$ & $0.838^{\star\star}$ & $0.916^{\star\star}$ & $0.61^{\mathrm{ns}}$ & $0.204^{\star}$ \\
    \textbf{Ours w/ DRS}  & $\mathbf{18.95}$ & $\mathbf{0.840}$ & $\mathbf{0.918}$ & $\mathbf{0.61}$ & $\mathbf{0.202}$ \\
    \bottomrule
\end{tabularx}
    \label{r_1}
\end{table}
\subsection{Registration Derived Regional Volume Change}
We used a multiresolution, mass-preserving DIR framework for estimating the B-Spline parameterized transformation from the moving image $\boldsymbol{I}_{\mathrm{TLC}}$ to the fixed image $\boldsymbol{I}_{\mathrm{RV}}$~\cite{cao2010regularized, yin2009mass}. The variational framework used to estimate this transformation $\varphi(\cdot)$ minimized a sum of squared tissue volume difference (SSTVD) cost to preserve tissue volume change between two images~\cite{yin2009mass}. The Jacobian of deformation field ($\boldsymbol{J}_{\mathrm{SSTVD}}$) was used to measure local tissue volume change (see Figure 1a). In this work, $\boldsymbol{J}_{\mathrm{SSTVD}}$ was used as the target image for training the proposed generative model. 

\subsection{Preprocessing}
$\boldsymbol{I}_{\mathrm{TLC}}$ and $\boldsymbol{I}_{\mathrm{RV}}$ were resampled to have isotropic voxels with size $1\times1\times1$ mm$^3$. The volumes were cropped to the lung region using lung segmentations obtained from a multi-resolution convolutional neural network~\cite{gerard2020multi}. To remove outliers arising due to calcification or metal artifacts, CT intensity values were restricted from $-1024$ Hounsfield units (HU) to $1024$ HU. Similarly, intensity values of $\boldsymbol{J}_{\mathrm{SSTVD}}$ were clipped between the interval $[\mu - 3\sigma, \mu + 3\sigma]$, where $\mu = \sum^{N}_{i = 1} \mathrm{mean} (\boldsymbol{J}_{i}) / N$ and $\sigma = \sum^{N}_{i = 1} \mathrm{std} (\boldsymbol{J}_{i}) / N$ for the training set of size $N$. Next,  the image intensities were rescaled to the interval $[-1, 1]$ for training.
We trained our models using mid-coronal slices from each volume, which were padded or cropped to a fixed size of $256 \times 256$ for constructing reasonable batch sizes. DIR of $\boldsymbol{I}_{\mathrm{TLC}}$ and $\boldsymbol{I}_{\mathrm{RV}}$ yielded $2500$ $\boldsymbol{I}_{\mathrm{RV}}$ and $\boldsymbol{J}_{\mathrm{SSTVD}}$ pairs, $1761$ of which were used for training and $739$ for evaluation, as shown in Table \ref{t_1}. The training data had $311,296$ slices and the testing data had $130,516$ slices.

\begin{figure*}[t!]
\centering
\includegraphics[scale = 0.33]{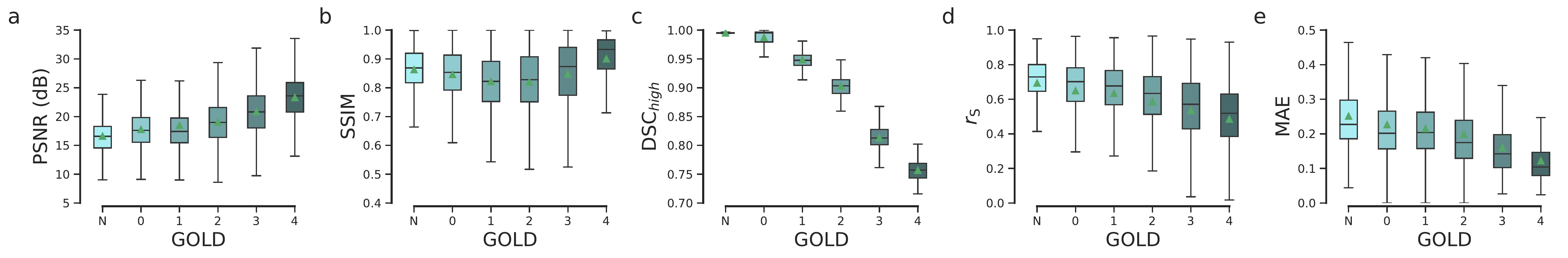}
\caption{Performance of proposed framework with DRS across varying GOLD stages. (Horizontal lines indicate medians while triangles indicate means.)}
\label{metrics_by_gold}
\end{figure*}

\subsection{The Proposed Method}
\subsubsection{Conditional Least Squares GAN (cLSGAN)}
We developed an image-conditional generator $\mathcal{G}$ to learn a map $\mathcal{G}: \mathbf{x} \to \mathbf{y}$, where $\mathbf{x},\: \mathbf{y} \in \mathbb{R}^{256 \times 256}$ were slices from $\boldsymbol{I}_{\mathrm{RV}}$ and $\boldsymbol{J}_{\mathrm{GAN}}$ volumes, respectively (see Figure 1b). The generated slices $\hat{\mathbf{y}} = \mathcal{G}(\mathbf{x})$ were then evaluated by the discriminator model $\mathcal{D}$, which is typically a classifier that minimizes cross-entropy loss~\cite{goodfellow2014generative}. Using cross-entropy as an objective, however, is known to cause training instability due to vanishing gradients. Therefore, we replaced it with least squares adversarial loss to address training instability~\cite{mao2017least}. The least squares GAN (LSGAN) objective for an image-conditional framework could thus be expressed as:
\begin{equation}
\label{cgan}
    \mathcal{L}_{\mathrm{cLSGAN}} = -\:\mathbb{E}_{\mathbf{x},\: \mathbf{y}}[(\mathcal{D}(\mathbf{x}, \mathbf{y}) - 1)^{2}] - \mathbb{E}_{\mathbf{x}}[\mathcal{D}(\mathbf{x}, \mathcal{G}(\mathbf{x}))^{2}].    
\end{equation}

In addition to the adversarial feedback provided to generator by discriminator, various image-to-image translation methods (e.g., Pix2Pix~\cite{isola2017image}) minimize $L_{1}$ or $L_{2}$  distance  between the target $\mathbf{y}$ and generated samples $\hat{\mathbf{y}}$. Although $L_{2}$ loss is convex it is sensitive to outliers, while $L_{1}$ loss is robust to outliers but non-convex. Alternatively, we used the Charbonnier or pseudo-Huber loss which is strongly convex around zero and robust to outliers as well~\cite{barron2019general}. The overall loss function thus minimized by the generator $\mathcal{G}$ was:
\begin{equation}
    \mathcal{L}_{\mathcal{G}}(\mathbf{x},\: \mathbf{y}) = \underbrace{\mathcal{L}_{\mathrm{MSE}}(\mathcal{D}(\mathbf{x},\: \mathcal{G}(\mathbf{x})), 1)}_\text{adversarial loss~} + \underbrace{\lambda \mathcal{L}_{\mathrm{Ch}}(\mathbf{x},\: \mathbf{y})}_\text{Charbonnier loss~},
\end{equation}
where $\mathcal{L}_{\mathrm{Ch}} = \mathbb{E}_{\mathbf{x},\: \mathbf{y}}[\rho(\mathbf{y} - \mathcal{G}(\mathbf{x}))]$, $\rho(t) = \sqrt{t^{2} + \epsilon^{2}}$, $\lambda = 200$, and $\epsilon$ was a very small constant set to $10^{-6}$. For the generator $\mathcal{G}$, we used a UNet-like encoder-decoder architecture with skip-connections~\cite{isola2017image}. Instead of a conventional classifier, we used the PatchGAN~\cite{isola2017image} discriminator $\mathcal{D}$ for eliciting patchwise feedback on real and synthetic samples.  Both networks were trained using the Adam optimizer with imbalanced learning rates -- $0.0002$ for generator and $0.0001$ for discriminator. 

\vspace{-0.1in}
\subsubsection{Style Transfer via Instance Normalization}
Instance normalization (IN) was recently shown to be an efficient component for fast neural style transfer~\cite{ulyanov2016instance}, and performed better than the commonly used batch normalization (BN). To encourage better style transfer, we replaced BN from each encoder-decoder block with IN. Thus, each block of encoder-decoder was comprised by $\mathrm{2D}$  convolution $\rightarrow$ IN $\rightarrow$ ReLU (rectified linear unit).

\subsubsection{Discriminator Rejection Sampling (DRS)}
A recently identified property of the discriminator $\mathcal{D}$ is mode preservation -- as it could be used to prevent the generator $\mathcal{G}$ from moving to irrelevant modes~\cite{azadi2018discriminator}. The samples evaluated to be least realistic by a sufficiently trained $\mathcal{D}$ were shown to be responsible for errors in generalization~\cite{azadi2018discriminator}. Therefore, we used a simplified implementation of DRS and trained our network by updating gradients only from $k$ batch-samples with the highest score by $\mathcal{D}$~\cite{sinha2020top}. Initially set as $k = B$, $k$ was decayed every $10$ epochs ultimately to $k = B / 2$, where batch size $B = 64$.

\begin{figure}[h!]
\centering
\includegraphics[scale = 0.22]{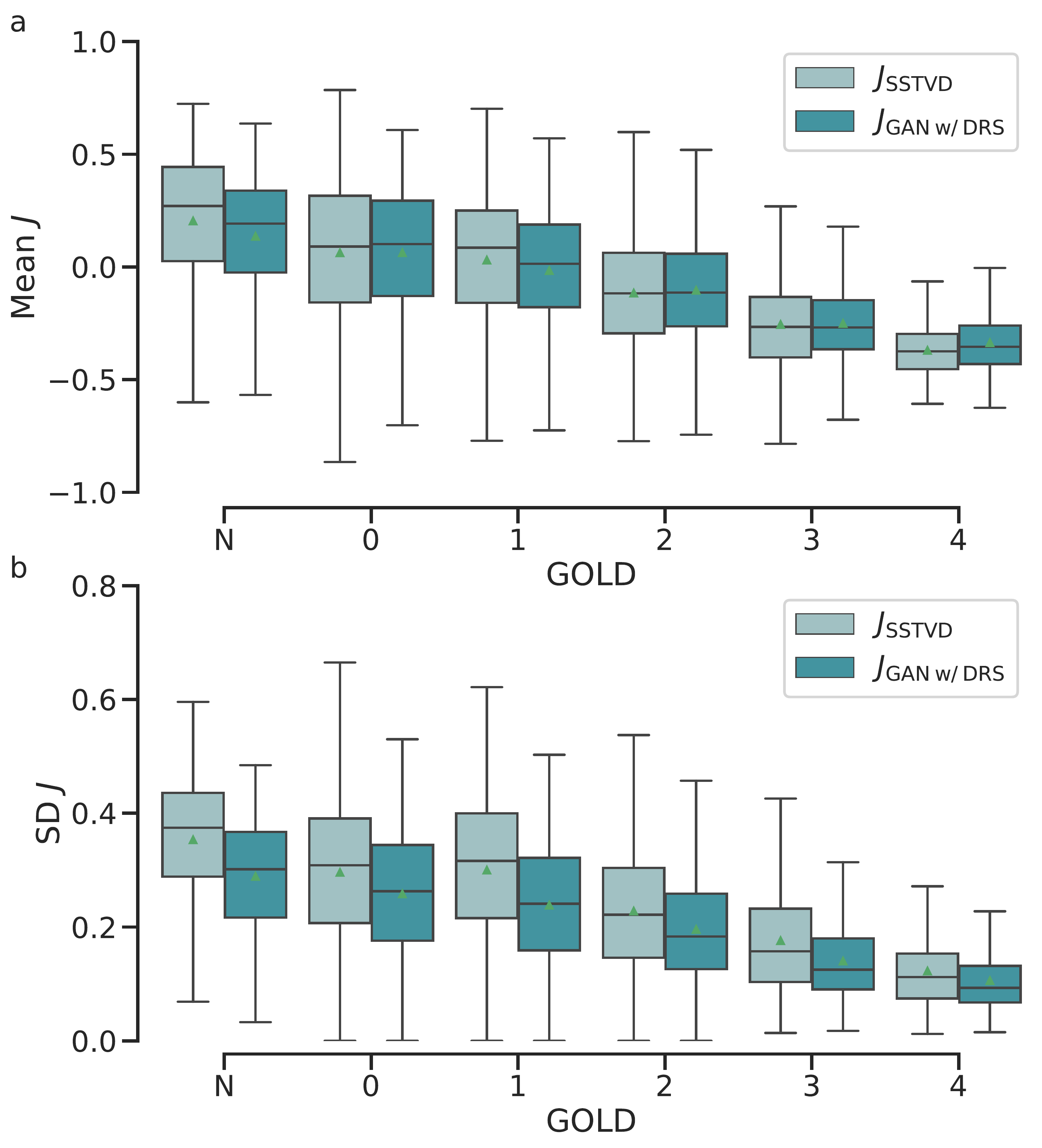}
\caption{Distribution of global image statistics (mean and SD) from DIR-based $\boldsymbol{J}_{\mathrm{SSTVD}}$ and single-volume $\boldsymbol{J}_{\mathrm{GAN\:w/\:DRS}}$ across different GOLD stages (horizontal lines indicate medians while  triangles denote means).}
\label{global_stats}
\end{figure}

\subsection{Performance Evaluation}
We used several local and global metrics for assessing the quantitative performance of our framework. To evaluate voxelwise coherence, we assesed the peak signal to noise ratio (PSNR) and mean absolute error (MAE) between $\boldsymbol{J}_{\mathrm{SSTVD}}$ and $\boldsymbol{J}_{\mathrm{GAN}}$. To evaluate structural consistency, the structural similarity index (SSIM) was used~\cite{wang2004image}. We used the Dice similarity coefficient (DSC) of high-functioning regions, defined as pixels above $75^{th}$ percentile, between $\boldsymbol{J}_{\mathrm{SSTVD}}$ and $\boldsymbol{J}_{\mathrm{GAN}}$ to ensure they were captured adequately. Overall voxelwise correlation was also evaluated using Spearman's correlation $r_{s}$. We also compared distributions of global statistical measures, including mean and standard deviation (SD), from DIR-derived $\boldsymbol{J}_{\mathrm{SSTVD}}$ and generated $\boldsymbol{J}_{\mathrm{GAN}}$  tissue volume expansion slices. The performance of our framework was compared with a UNet based non-adversarial network that minimized SSIM- and $L_{1}$- loss~\cite{gerard2020estimating}, and another well-known image-to-image translation framework -- Pix2Pix~\cite{isola2017image}.

\section{Results}
\label{sec:results}
\begin{figure}[ht!]
\centering
\includegraphics[scale = 0.155]{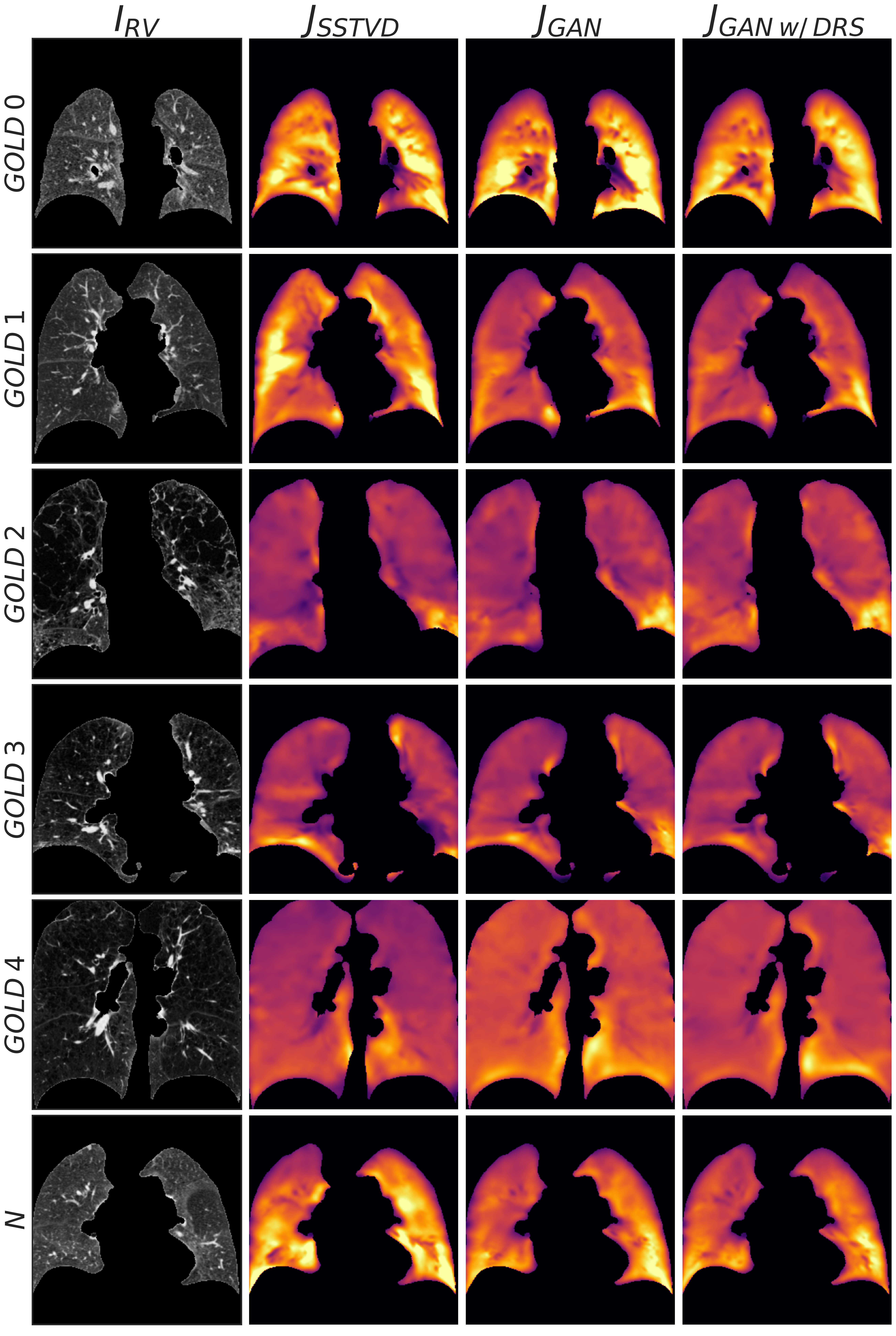}
\caption{Qualitative synthesis results from the proposed framework ranging from GOLD 0 to GOLD 4, and non-smokers (N). $\boldsymbol{J}_{\mathrm{SSTVD}}$ represents two-volume DIR approach, while $\boldsymbol{J}_{\mathrm{GAN}}$ and $\boldsymbol{J}_{\mathrm{GAN \: w/ \:DRS}}$, represent tissue expansion computed by the proposed single-volume GAN approach.}
\label{j_with_gold}
\end{figure}

Quantitative evaluation for all models is shown in Table~\ref{r_1}. The proposed model achieved higher PSNR, SSIM, $r_s$, and MAE and the differences were significant ($p < 0.0001$), as suggested by Mann Whitney's U test. Model performance by GOLD stage is given in Figure~\ref{metrics_by_gold}. Global statistics of $\boldsymbol{J}$ across GOLD stages are shown in Figure~\ref{global_stats}. Representative samples across different GOLD stages are shown in Figure~\ref{j_with_gold}. The sagittal plane of an example case is shown Figure~\ref{ventral-dorsal}, this was generated by independent model predictions on all coronal slices.

\vspace{-0.1in}
\section{Discussion and Conclusion}
\label{sec:discussion}
A generative adversarial learning framework was proposed to estimate local tissue expansion from  CT at single expiratory volume. The method was evaluated on a large cohort of $739$ subjects from different COPD GOLD stages and performed well across five quantitative metrics. Model performance was consistent for different GOLD stages, except DSC$_{\mathrm{high}}$ which decreased due to an imbalanced representation of high-functioning regions across later disease stages. The proposed method was able to learn variations in COPD severity, as means and SDs of $\boldsymbol{J}_{\mathrm{SSTVD}}$ and $\boldsymbol{J}_{\mathrm{GAN}}$ decayed with GOLD stage, corroborating the clinical findings by Bhatt \textit{et al}.~\cite{bhatt2017computed} Recently, a non-adversarial generative framework was trained on inspiratory-expiratory volumes pairs to automatically estimate local tissue expansion~\cite{gerard2020estimating}. The study proposed to minimize SSIM-loss but still utilized dual-volume scans for estimating tissue expansion which was computed at a significantly lower resolution. 

\begin{figure}[t!]
\centering
\includegraphics[scale = 0.32]{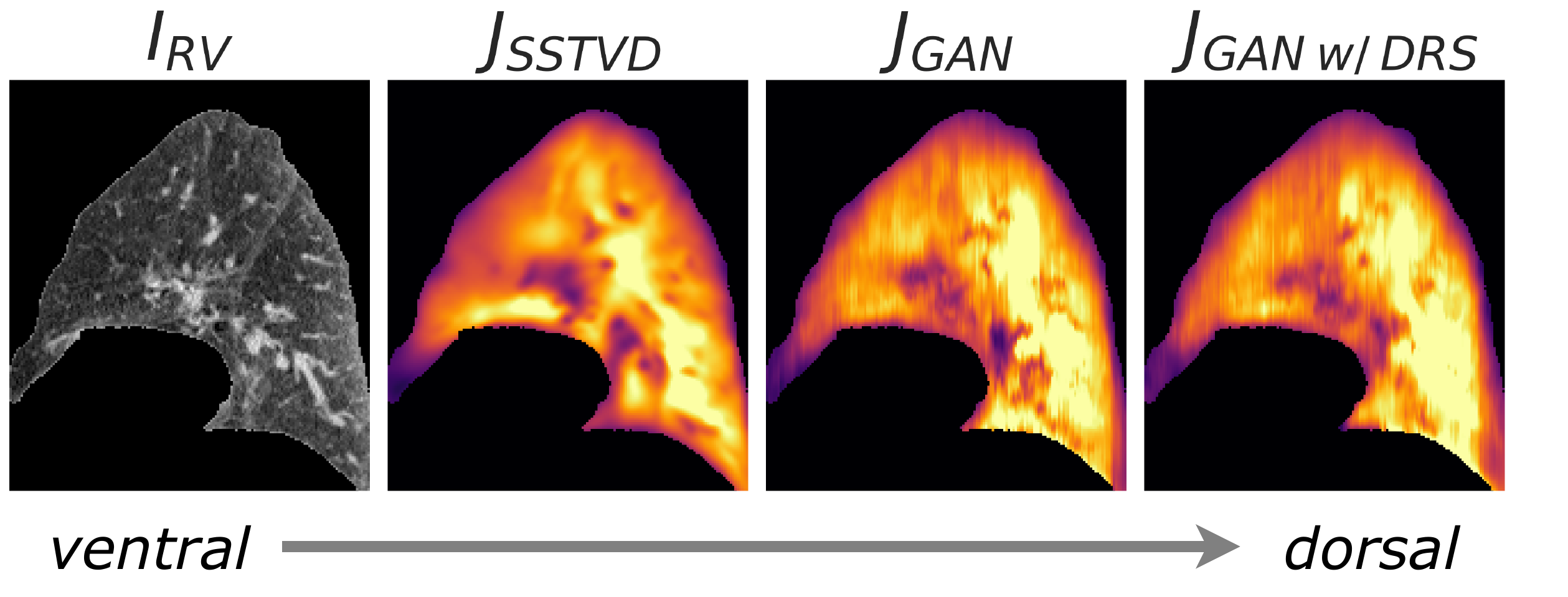}
\caption{Ventral to dorsal gradient patterns shown on sagittal slices from $\boldsymbol{J}_{\mathrm{SSTVD}}$, $\boldsymbol{J}_{\mathrm{GAN}}$, and $\boldsymbol{J}_{\mathrm{GAN \: w/ \:DRS}}$.}
\label{ventral-dorsal}
\end{figure}

A limitation of the proposed method is that it was trained on 2D coronal slices, which could lead to discontinuities between slices and prevent the network from learning ventral-dorsal patterns. However, the sagittal slices were visually inspected and we found that the network appropriately captured the ventral-dorsal gradients observed on supine scans due to gravity (Figure~\ref{ventral-dorsal}). Another limitation of our framework is its potential sensitivity to the underlying DIR method, which needs to be investigated across different registration methods.

The trained generator can be used as a registration-independent tool for estimating local tissue mechanics. By eliminating the need for multiple breath-hold scans, the approach can significantly reduce exposure to radiation dose, and can be applied to broader patient strata. 

\section{Compliance with Ethical Standards}
This research study was conducted retrospectively using anonymized human subject data made publicly available by SPIROMICS. Written consent was provided by all subjects, and the protocols were approved by the Institutional Review Boards (IRBs) of each participating study center.

\section{Acknowledgments}
\label{sec:acknowledgments}
This work was supported in part by the grants R01HL142625 and by a grant from The Roy J. Carver Charitable Trust. SPIROMICS was supported by contracts from the NIH/NHL BI (HHSN268200900013C, HHSN268200900014C, HHSN 268200900015C, HHSN268200900016C, HHSN268200900 017C, HHSN268200900018C, HHSN268200900019C, HHS N268200900020C), grants from the NIH/NHLBI (U01 HL1 37880 and U24 HL141762), and supplemented by contributions made through the Foundation for the NIH and the COPD Foundation from AstraZeneca/MedImmune; Bayer; Bellerophon Therapeutics; Boehringer-Ingelheim Pharmaceuticals, Inc.; Chiesi Farmaceutici S.p.A.; Forest Research Institute, Inc.; GlaxoSmithKline; Grifols Therapeutics, Inc.; Ikaria, Inc.; Novartis Pharmaceuticals Corporation; Nycomed GmbH; ProterixBio; Regeneron Pharmaceuticals, Inc.; Sanofi; Sunovion; Takeda Pharmaceutical Company; and Theravance Biopharma and Mylan.

\bibliographystyle{ieeetr}
\bibliography{References}

\end{document}